# Factors Influencing Job Rejections in Cloud Environment

**Rashmi K S, Suma. V & Vaidehi. M**

Dayananda Sagar College of Engineering, Bangalore, India
E-mail : rashmiks.ks57@gmail.com , sumavdsce@gmail.com, dm.vaidehi@gmail.com

**Abstract -** The IT organizations invests heavy capital by consuming large scale infrastructure and advanced operating platforms. The advances in technology has resulted in emergence of cloud computing, which is promising technology to achieve the aforementioned objective. At the peak hours, the jobs arriving to the cloud system are normally high demanding efficient execution and dispatch. An observation that has been carried out in this paper by capturing a job arriving pattern from a monitoring system explains that most of the jobs get rejected because of lack of efficient technology. The job rejections can be controlled by certain factors such as job scheduling and load balancing. Therefore, in this paper the efficiency of Round Robin (RR) scheduling strategy used for job scheduling and Shortest Job First Scheduling (SJFS) technique used for load balancing in reducing the job rejections are analyzed. Further, a proposal for an effective load balancing approach to avoid deadlocks has been discussed.

## I. INTRODUCTION

The IT organizations are oriented towards the development of cost, time and resource effective products. Cloud computing is one of the promising technology to achieve the aforementioned objective. Further, cloud computing is a pay-go-model which provides on demand network access to a shared pool of computing resources that can be rapidly provisioned and released with minimal management to its clients as a metered service. But in an IT organization jobs arriving to the system are normally high at peak hours, which demands efficient execution. Further, virtualization is the core characteristic of cloud computing, where the jobs that arrive to the system are diverse in nature and runs independently on the virtualized operating system. Consequently, to satisfy the request, the resources have to be properly utilized such that there is no deterioration in the working of the system which in turn leads to high rejection of the jobs. If the job rejection rate is high then the business of the cloud service provider also deteriorates. As cloud is a pay-go-model, the business performance needs to be accelerated which is a challenging issue in the domain. There are certain important factors namely job scheduling, load balancing and resource allocation in the cloud environment that supports to improve the business performance by reducing the job rejections.

An efficient job scheduling strategy must aim to yield less response time so that the execution of submitted jobs takes place within a stipulated time and simultaneously there will be an occurrence of in-time resource reallocation. As a result of this, less rejection of jobs takes place and more number of jobs can be submitted to the cloud by the clients which ultimately results in accelerating the business performance of the cloud system.

As the resources are provided on demand in the cloud system there is a necessary for the resources to be highly available. Failure of which, results in job starvation, which is a situation where a job does not get the resource that it requires for a long time because the resource is being allocated to other jobs. Hence, resource management is a complex task from the business point of the cloud service provider. Henceforth, an effective load balancing strategy must provide a solution to the aforementioned issue by yielding less response time which in turn results in less starvation and reduced job rejections which are the necessary factors to accelerate the business of the cloud service provider. However, in the cloud environment there are possibilities of deadlock occurrences when there is more than one request arriving to the system for competing to acquire the same resource at the same instance of time. Therefore, an effective load balancing strategy must be incorporated to avoid deadlocks.

Henceforth, in this paper a survey on the factors that governs to accelerate the business performance of the cloud has been provided.

The paper is organized as follows. The Section II of this paper explains the related work, Section III elucidates the research design and Section IV explains the research methodology. Finally, the paper is concluded in Section V followed by references.

## II. RELATED WORK

The emergence of cloud computing in the IT industry has opened several research avenues in the domain. Authors of [1] have proposed an improved activity-based costing [IABC] algorithm for task scheduling in cloud computing with the objective to schedule group of tasks in cloud computing platform with resources having different costs and computation performance. The result of the algorithm shows that the time taken to complete the tasks is less by grouping them than without grouping them. However, this algorithm cannot handle simultaneous tasks.

However, authors in [2] have suggested a task scheduling technique using credit based assignment problem in cloud computing to find minimal completion time of tasks. They have considered a single parameter i.e., cost to minimize the completion time of tasks. However, in the cloud environment it is necessary to achieve optimal solution. Henceforth, it is imperative to consider all parameters which influence the realization of minimal completion time of tasks.

Authors in [3] have implemented efficient resource provisioning in compute clouds through VM multiplexing. In this paper, resources provisioning is based on the estimate of the capacity needs of the virtual machines. Further, the virtual machines are multiplexed using VM selection method, which uses the correlation among the VMs as an indicator of their compatibility. Hence, with this approach high utilization of the resources is achieved. But, the limitation of this approach is that there is an overhead involved during the VM selection process.

Authors in [4] have proposed a load balancing strategy for virtual storage to provide a large scale net data storage model and Storage as a Service model based on cloud. Further the storage virtualization is achieved using three layers architecture with two load balancing modules to balance the load. The strategy implemented in this paper is limited to the cloud service providers providing Storage as a Service (SaaS).

Authors in [5] have recommended load balancing in a three–level cloud computing network, by using a scheduling algorithm which combines the features of Opportunistic Load Balancing (OLB) and Load Balance Min-Min (LBMM) which can utilize better executing efficiency and maintain load balancing of the system. The objective is to select a node based for executing the complicated tasks that needs large-scale computation. The scheduling algorithm proposed in this paper is not dynamic and also there is an overhead involved in the selection of the node.

## III. RESEARCH DESIGN

An effective resource model is very much necessary as it impacts the performance of the cloud computing [14]. As cloud is a pay-go model, it is necessary to have an efficient task scheduling strategy and an effective load balancing strategy to accelerate the business performance of the cloud system. Hence, it is essential that the resource model should support optimized utilization of available resources and execution of tasks within a stipulated time in the cloud environment.

In this investigation, a secondary data has been collected for analyzing the efficiency of the cloud service provider for in time execution of jobs at the peak hours with available resources. The secondary data is the processed data collected from several IT industries functioning in the cloud environment.

It is apparent from the analysis that task scheduling and load balancing strategies are the influencing factors in cloud to reduce job rejections and there by accelerates the business performance of the cloud system.

## IV. RESEARCH METHOD

Figure 1. is the flow diagram of this investigation depicting the job processing in cloud . In this diagram diverse requests arrive from various users to the cloud datacenter. The cloud datacenter process these requests and dispatches them. The processing of request will be either through the job scheduling or through the load balancing. The existing job scheduling strategy operates in Round Robin (RR) fashion to yield less response time for reducing the job rejections at the peak hours while Shortest Job First Scheduling (SJFS) technique is used for load balancing, also yields less response time to reduce job rejections as well as to reduce starvations. Further, the load balancing should avoid deadlocks by promoting Virtual Machine (VM) migrations which is another important factor to reduce the job rejections. Subsequently, the reduction in job rejection ultimately results in accelerating the business performance of the cloud system

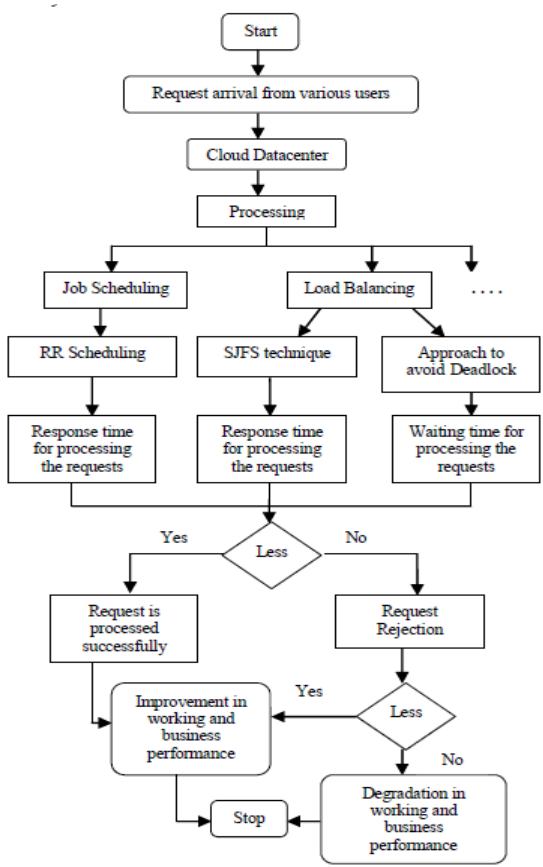

Fig. 1 : Job Processing in Cloud

The existing job scheduling technique in cloud is the Round Robin (RR) scheduling technique. Henceforth, in order to analyze the efficiency of RR scheduling technique a simulation is carried out using CloudAnalyst simulator. TABLE I., TABLE II. and TABLE III. depicts the simulation configuration.

TABLE I. depicts the details of user base configuration, which provides details related to request for every hour, data size for every request and datacenter where the requests are processed.

### TABLE I : USER BASE CONFIGURATION

| UB | R/U/H | DS/R | DC |
|-----|-------|--------|------|
| UB1 | 12 | 100 | DC4 |
| UB2 | 12 | 10000 | DC3 |
| UB3 | 12 | 100000 | DC1 |
| UB4 | 12 | 1000 | DC2 |
| UB5 | 12 | 10000 | DC2 |

UB – User base, R/U/H –Request per user per hour, DS/R – Data Size per request, DC – Datacenter.

TABLE II. shows the Datacenter configuration, which provides details related to the number of VMs present in DC, memory and bandwidth capability of every VM.

### TABLE II : DATACENTER CONFIGURATION

| DC | No. of VMs | Mem. | BW |
|------|-----------|------|-------|
| DC1 | 40 | 1024 | 1000 |
| DC2 | 20 | 512 | 100 |
| DC3 | 50 | 512 | 10000 |
| DC4 | 35 | 1024 | 1000 |

DC – Datacenter, Mem – Memory, BW - Bandwidth

TABLE III. shows the information of advanced configuration, which comprises of user grouping factor in user base, request grouping factor in datacenters and executable instruction length for every request.

### TABLE III : ADVANCED CONFIGURATION

| User grouping factor | 1000 |
|-------------------------------|------|
| Request grouping factor | 100 |
| Executable instruction length | 250 |

The result of this simulation has been obtained in terms of response time and processing time and TABLE IV. depicts the result of this simulation.

### TABLE IV : RESPONSE TIME AND PROCESSING TIME OBTAINED USING RR SCHEDULER

| Mm | RT (ms) | DC PT (ms) |
|-----------|---------|------------|
| Avg. (ms) | 115.23 | 2.51 |
| Min. (ms) | 44.33 | 0.08 |
| Max. (ms) | 380.03 | 7.98 |

Mm – Measurement metric, RT –Response time, DC PT – Datacenter Processing Time

Figure 2. shows a sample graph of user base hourly average response time using the RR scheduler.

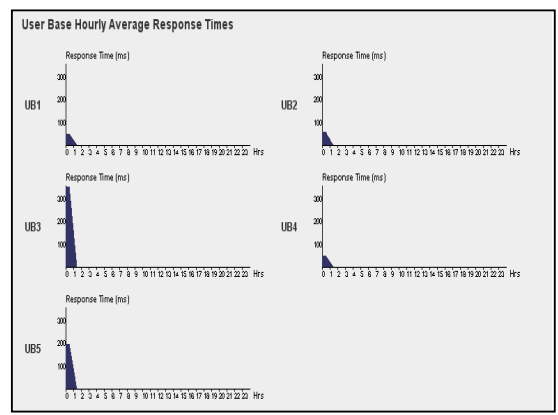

Fig. 2 : A Graph of User Base Hourly Average Response time using RR scheduler

From the above graph it is analysed that there is a high response time leading to very high proceesing time and ultimately resulting in increased rejection in the number of the jobs submitted. Figure. 3. illustrates the number of jobs rejected at the peak hours graphically. This figure further infers that with increrase in the submission rate of jobs, rejection rate further increases. As a supporting to the aforementioned TABLE V. gives the percentage of job rejected using RR scheduler.

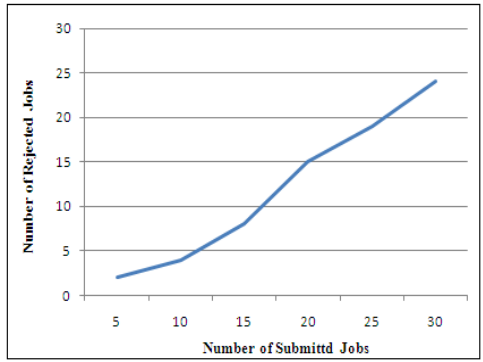

Fig. 3 : Graph Showing the number of jobs rejected using RR scheduler

TABLE V : PERCENTAGE OF JOB REJECTIONS USING RR SCHEDULER

| No. of Jobs Submitted | Percentage of Jobs Rejected |
|---|---|
| 5 | 40 |
| 10 | 40 |
| 15 | 53 |
| 20 | 75 |
| 25 | 76 |
| 30 | 80 |

As discussed earlier, the Load balancing is also one of important factor that influences the job rejection. The existing load balancing technique in cloud is Shortest Job First Scheduling (SJFS). This is further incorporated in the Load Balancer to analyze its efficiency in reducing the job rejections, in which the job with less burst time will be scheduled first by the load balancer.

TABLE VI. depicts the list of jobs received with their IDs and the burst time. According to the SJFS technique, the job with less burst time first gets scheduled for allocation. Therefore the drawback of this algorithm is that the jobs with higher burst time will be starved and gets rejected if the job is starving for a longer time and ultimately degrading the system performance.

TABLE VI : A SAMPLE OF JOBS RECEIVED

| Job ID | Arrival time (hrs) | Burst time (hrs) |
|---|---|---|
| 1 | 0 | 8 |
| 2 | 1 | 4 |
| 3 | 3 | 6 |
| 4 | 5 | 2 |
| 5 | 6 | 5 |

The scheduling order for the jobs shown in TABLE VI. will be Job 1, Job 4, Job 2, Job 5, Job 3 and the TALBE VII. depicts the Response time obtained from using the SJFS technique for Load Balancing and Figure 4. shows this graphically.

TABLE VII : RESPONSE TIME OBTAINED USING SJFS FOR LOAD BALANCING

| Job ID | Response Time (hrs) |
|---|---|
| 1 | 0 |
| 4 | 3 |
| 5 | 8 |
| 2 | 9 |
| 3 | 16 |

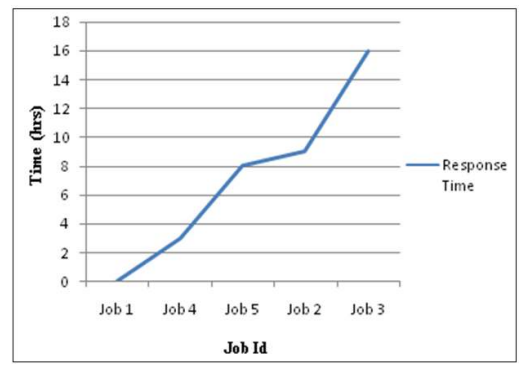

Fig. 4 : A Graph showing the Response Time obtained using SJFS technique for Load Balancing

From TABLE VII. and Figure. 4. it is analyzed that the Job 3 has got very high response time and it will be in starvation for a very long time resulting in rejection. Hence, with this strategy, increase in the number of job arriving to the system increases the response time and thereby results in starvation and job rejections. Figure 5. illustrates the number of jobs rejected at peak hours graphically by using the SJFS technique for load balancing. TABLE VIII. shows the percentage of job rejections using SJFS technique for load balancing. It is indicated from this table that with increased submission of jobs increases job rejections resulting in business loss to the cloud service provider.

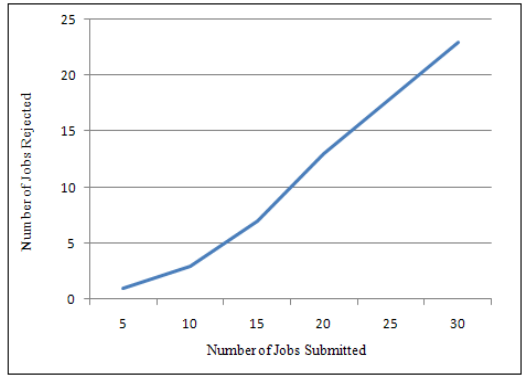

Fig. 5 : A Graph showing the number of jobs rejected using the SJFS technique for Load Balancing

TABLE VIII : PERCENTAGE OF JOBS REJECTED USING SJFS TECHNIQUE FOR LOAD BALANCING

| No. of Jobs Submitted | Percentage of Jobs Rejected |
|---|---|
| 5 | 20 |
| 10 | 30 |
| 15 | 47 |
| 20 | 65 |
| 25 | 72 |
| 30 | 77 |

Another important challenge to reduce job rejections in the cloud environment is to avoid deadlocks among the jobs which are contending for the same resource at the same instance of time [15]. This issue can be handled through load balancing by promoting VM migrations such that jobs will be migrated from the over loaded VM to the under-utilized VM by comparing the wait time and the hop time. Henceforth, the availability of resources increases and the deadlock is avoided. Currently, in the cloud computing domain there is no such load balancing approach that can avoid deadlocks, which reduces job rejection.

## V. CONCLUSION

Cloud Computing is a promising technology to support IT organizations in developing cost, time and resource efective products. Since, Cloud computing is a pay-go-model, it is necessary to reduce job rejections at the peak hors inoredr to improve the business performance of the cloud system. The job rejection can be reduced through job scheduling and load balancing tecjnique.

This investigation, analyzes the efficiency of Round Robin (RR) scheduling strategy used for job scheduling and Shortest Job First Scheduling (SJFS) technique used for load balancing to reduce the job rejections. It is evident from this analysis that both the strategies can be further enhanced inoder to reduce job rejections, which is necessary to accelerate the working and business performance of the clud system. Further, yo avoid deadlocks in the cloud system it is recommended to have an effective load blancing approach that can efficiently allocate the resources to resolve the dealdlocks.